\documentstyle{BSAXwork}

\input epsf.sty

\def\mincir{\ \raise -2.truept\hbox{\rlap{\hbox{$\sim$}}\raise5.truept	
\hbox{$<$}\ }}								%
\def\magcir{\ \raise -2.truept\hbox{\rlap{\hbox{$\sim$}}\raise5.truept	%
\hbox{$>$}\ }}								%

\def \AAP #1 #2 {{\em Astron. Astrophys.\/} {\bf #1}, #2}
\def \AAL #1 #2 {{\em Astron. Astrophys. Lett.\/} {\bf #1}, L#2}
\def \AAR #1 #2 {{\em Astron. Astrophys. Rev.\/} {\bf #1}, #2}
\def \AAS #1 #2 {{\em Astron. Astrophys. Suppl. Ser.\/} {\bf #1}, #2}
\def \AJ #1 #2 {{\em Astron. J.\/} {\bf #1}, #2}
\def \ANNREV #1 #2 {{\em Ann. Rev. Astron. Astrophys.\/} {\bf #1}, #2}
\def \APJ #1 #2 {{\em Astrophys. J.\/} {\bf #1}, #2}
\def \APJL #1 #2 {{\em Astrophys. J. Lett.\/} {\bf #1}, L#2}
\def \APJS #1 #2 {{\em Astrophys. J. Suppl.\/} {\bf #1}, #2}
\def \APSS #1 #2 {{\em Astrophys. Space Sci.\/} {\bf #1}, #2}
\def \ASR #1 #2 {{\em Adv. Space Res.\/} {\bf #1}, #2}
\def \BAIC #1 #2 {{\em Bull. Astron. Inst. Czechosl.\/} {\bf #1}, #2}
\def \JSQRT #1 #2 {{\em J. Quant. Spectrosc. Radiat. Transfer\/} {\bf #1}, #2}
\def \MN #1 #2 {{\em Mon. Not. R. Astr. Soc.\/} {\bf #1}, #2}
\def \MEM #1 #2 {{\em Mem. R. Astr. Soc.\/} {\bf #1}, #2}
\def \PLR #1 #2 {{\em Phys. Lett. Rev.\/} {\bf #1}, #2}
\def \PASJ #1 #2 {{\em Publ. Astron. Soc. Japan\/} {\bf #1}, #2}
\def \PASP #1 #2 {{\em Publ. Astr. Soc. Pacific\/} {\bf #1}, #2}
\def \NAT #1 #2 {{\em Nature\/} {\bf #1}, #2}
\def \SAIT #1 #2 {{\em Mem.\ Soc.\ Astron.\ It.\/} {\bf #1}, #2}
\def \MESS #1 #2 {{\em The Messenger\/} {\bf #1}, #2}
\def \ASTRNACH #1 #2 {{\em Astron. Nach.\/} {\bf #1}, #2}

\begin{opening}

\title{Revisiting the Blazar Main Sequence} 
\author{A. Cavaliere$^{1}$, V. D'Elia $^2$}
\institute{$^1$Dept. of Physics, Univ. of Rome Tor Vergata, Italy\\
$^2$Rome Observatory, Monteporzio, Italy}
\date{} 
\end{opening}

\begin{document}

\oddpagefooter{}{}{} 
\evenpagefooter{}{}{} 
\medskip  

\begin{abstract} 
A discussion  of the FSRQ -- BL Lac unification. All distinctive features 
marking these two Blazar subclasses find an unifying explanation if the sources 
are powered by central engines constituted by similar Kerr holes,  
but fueled at high and low  accretion rates, 
respectively. The connection need not be 
a genetic one, but evidence toward some FSRQs switching into BL Lacs at 
lower $z$  will be provided by moderately negative BL Lac evolution. Then 
an extrapolation will be warranted toward ultra-high energy 
particle accelerators operating at very low accretion rates.    
\end{abstract}

\medskip

\section{Introduction}
The BL Lac Objects may be described  as {\it minimal} Blazars. 
This is because   
-- unlike  the FSRQs which constitute the other main Blazar subclass --      
they feature at the same time  weak or no emission lines and 
blue-UV bumps, weak or moderate intrinsic power, and little or no 
cosmological evolution (for general descriptions and detailed references, 
see Urry \& Padovani 1995, Padovani \& Urry 2001). Being so rare, elusive, 
with redshifts not easily pinpointed, and 
hence so vulnerable to selection effects, 
these sources make difficult observational targets.

Yet the  BL Lacs constitute the focus of keen and mounting interest,  
being {\it extreme} Blazars that shine by nearly pure non-thermal 
radiation with little or no photon reprocessing. 
Their  spectra are exceedingly wide, in that they cover 
16 frequency decades or more, from the radio band (as also the FSRQs do) 
up to the high energy $\gamma$-rays beyond 10 GeV.  
In fact, a number of such objects have been observed into the 
TeV range (see Costamante et al. 2002). 

Concerning the spectral energy distribution 
(SED), all Blazars follow a roughly similar 
pattern  with two bumps mainly or mostly contributed by Synchrotron and Inverse 
Compton emissions. 
But (as pointed out by Fossati et al. 1998) in the  
FSRQs the frequency $\nu_p$ of the lower  peak is 
placed in the bands from the far IR to the optical, and the upper peak is 
around $10^2$ MeV. In the BL Lacs, instead, $\nu_p$ ranges from 
optical to X-rays, and the upper peak (possibly modified by  
the Klein-Nishina limit) falls around or above 10 GeV. 

In all Blazars such non-thermal emissions are sharply beamed; they  
arise from relativistic jets of particles 
with bulk Lorentz factor $\Gamma$ around 10 (see Sikora 2001), not 
substantially different for BL Lacs and FSRQs.  
These jets are seen nearly pole on within  angles of order  $ 1/\Gamma$,
with apparent powers enhanced by a ``blazing'' factor 
$\Gamma^2$ relative to the emitted ones (see Sikora et al. 1997, 
Ghisellini 1999).
What differs between the two Blazar subclasses is the 
total output produced; the top outputs are 
apparently limited to $L \sim 10^{46}$ erg/s in the BL Lacs, but 
may be up to $10^2$ times stronger in some FSRQs  
(see Maraschi 2001).
 
How and why all these BL Lac peculiarities arise together? 
Largely on {\it model-indepen-}  
{\it dent} 
grounds, we trace back their common origin  
to different fueling levels of basically similar power sources.

\section{The engine}  

The paradigm for the energetic cores of all AGNs is of course provided by 
accreting black holes with masses $M \sim 10^{8\pm1} \; M_{\odot}$.  
The accreting gas gathers in a disk where  its angular momentum is 
transferred outwards;  
in the process thermal, roughly isotropic emissions are generated with power 
$L_{th} \sim \eta\, L_E\, \dot m$ in terms of the 
conversion efficiency $\eta \sim 10^{-1}$, of the Eddington luminosity $L_E$ 
and of the corresponding  accretion rate $\dot m$. 
Does  the latter differ markedly in BL Lacs and in  FSRQs?  

An affirmative answer is provided 
evidence concerning the thermal emissions. 
The FSRQs, like the radioquiet QSs, emit   
a powerful ``big blue bump" conceivably from the inner disk, in addition to 
strong lines from continuum reprocessed by distant gas ``clouds"; 
the luminosities are high 
(Tavecchio et al. 2000) and 
close to the Eddigton values for BH masses around $10^9 \, M_{\odot}$, so    
$\dot m \sim 1 - 10$ is to hold.

In BL Lacs, instead, the absence or weakness of
both the emission lines ($EW \mincir 5$ \AA) and the 
BBB call for conditions of {\it low}  particle 
densities both around and in the accretion disks. 
 Since the latter  density scales as $n \propto \dot m^{11/20}$ and the 
former as $n \propto \dot m$ (see 
Frank, King \& Raine 2002), these conditions indicate
$\dot m \ll 1$. 
A similar conclusion holds if the BBB is due to 
emission from a hot corona (Sun \& Malkan 1989). 

On the other hand, evidence from the jets and their emission 
indicates the mass is not the only parameter 
relevant to the central BH in Blazars. Such an evidence  is provided, e.g.,   
by the  strong radio emission powered by jets which  has been recently
shown not to be simply  correlated with large host and BH masses (Ho 2002, 
McLure \& Dunlop 2002).  
So a further variable must enter;  we entertain the view 
involving the other basic parameter of the BHs, i.e., the angular momentum 
{\bf J} (see Blandford 1990, 1993). 

By its vectorial nature this is very likely to  provide   
the sharp and steady directionality of the jets.    
In addition, rotating Kerr holes 
make closer stable orbits (approaching  $r_g \approx GM/c^2 \approx 1.5\, 
10^{14} \,M_9$ cm) available to the gas particles before they plunge 
into the horizon; 
so more gravitational energy can be extracted, and the 
the maximal efficiency is raised from $\eta \sim 6 \, 10^{-2}$ to nearly $ 0.4$.
Finally, the rotational energy of a Kerr hole (proportional to 
$J^2$ in a classical rendition) 
may contribute directly the jet-like output   
of BL Lacs in particular. 

Part of the rotational energy of the hole can be directly extracted  
via hydromagnetic Poynting-like flux related 
to  {\bf B} and {\bf E} fields coherent on large scales, in  
approximately force-free conditions {\bf E} $\bullet$ {\bf B} $\approx 0$;
this is of course the attractive but debated scheme proposed by 
Blandford \& Znajek (1977), with the power produced $L_K$ ultimately 
depending on the 
square of the ${\bf B}$ field rooted in the disk but 
threading the horizon close to $r_g$.  
A similar hydromagnetic mode of energy extraction has been extended by 
Blandford \& Payne (1982) to the contribution $L_d$
from  the  accretion disk itself.

Thus two  contributions to the non-thermal 
power  $L_K + L_d = L_{nth}$ may be 
envisaged from the central region surrounding 
the Kerr hole within  radii  
$r_{e}\magcir r_g $, in the presence of  the poloidal field $B$.  
Both components scale approximately as     
%
%
$$L_{K} \propto B^2  r_{e}^2 ~, \eqno(1)$$
following the basic dependences provided by electrodynamics. 
But  the pressure $P = B^2/8 \pi$ holding 
the magnetic field, and the specific value of the   
effective radius $r_{e}$ are provided  by the disk structure, and  
this depends on the accretion rate $\dot m$. 

\section{Fueling rates}  

So both outputs $L_{nth}$ and 
$L_{th}$ are ultimately governed by the fueling regime of 
similar basic engines. 
But the latter output     
scales down as $L_{th} \propto \dot m$ when  $\dot m$ is decreased, 
while the same may not be the case with the former.   

Dependences and numerical coefficients in  $L_{nth}$  
have been focused and discussed  by Ghosh \& Abramowicz (1997); 
Moderski, Sikora \& Lasota (1998);  Livio, Ogilvie \& Pringle (1999).  
In the framework of standard disk models 
 the BZ77 scheme is found to yield the  maximal power  
%
$$L_{K} \approx 2\,10^{45}\,M_9\,(J/J_{M})^2\; erg\,s^{-1} ~,
\eqno (2)$$%
where $J_{M} 
\approx GM^2/c$ is the maximum value for the
angular momentum consistent with the Kerr metric. 
This value of $L_K$ is attained for accretion rates   
$\dot m  \magcir  10^{-2}$ when the inner disk is 
dominated by the radiation pressure  $P \propto M^{-1}$ 
independent of $\dot m$.

For lower $\dot m$ instead, the gas pressure $ P \propto M^{-9/10}
\, \dot m^{4/5}$ 
dominates also the inner  disk; then  
lower powers 
$L_{K}\approx 10^{44} \,M_9^{11/10}\, \dot m_{-4}^{4/5}\,(J/J_M)^2$ erg/s 
obtain. Yet the the ratio $L_{K}/L_{th}$  
exceeds unity and scales inversely with $\dot m$, as $ \dot m^{-1/5}$. 
The scaling is stronger if indeed at 
$\dot m \approx 10^{-2}$ 
a silent ADAF regime sets in (see Narayan, Mahadevan \& Quataert 1998; 
Frank et al. 2002); then the ions cannot share much energy with electrons 
before plunging into the horizon, 
and most radiations are suppressed sharply.  

We stress two main points of interest in these results. 
First, the emission becomes increasingly non-thermal anyway  
as $\dot m$ decreases below some $10^{-2}$. 
Second, for values $\dot m > 10^{-2}$ the power  saturates to  
$L_{K} \mincir 6 \, 10^{45}$ erg/s  
for $M \mincir 3 \; 10^9 \;M_{\odot}$. 
To this value one should add the 
the nonthermal disk contribution $L_d$, which in conditions of maximal $J$ 
and closest orbits with small $r_e \approx  r_g$ may be down to 
the same order as, or just a few times larger than $L_K$ (Meier 2001).   
Then the Kerr hole and inner disk may be considered as a dynamically 
and magnetically coupled system, using up also energy stockpiled  
in previous accretion episodes of mass and associated 
angular momentum (Bardeen 1970). 

Thus  a total $L_{nth} \approx  10^{46}$ erg/s appears to provide 
an upper {\it bound} to outputs mainly constituted by  
{\it non-thermal} radiation, the defining feature of a BL Lac;  
the condition  $L_{nth}/L_{th} \magcir 1$ remains relevant in spite of    
the blazing effects of the jets that tend to swamp a comparable thermal 
emission. 
This leads us to  propose  that top BL Lac outputs 
remaining below this limit provide evidence  that     
{\it rotational} energy is in fact extracted from the hole via  
the BZ77 mechanism.  The existing data 
(Maraschi 2001, see also fig. 1) support to now this prediction. 

The FSRQs, on the other hand, do have larger outputs that in some cases  
exceed $10^{47}$ erg/s
if protons contribute  more then the electrons
to the beam kinetic power (Celotti, Ghisellini \& Padovani 1997; 
Tavecchio et al. 2000).  
Within the above framework, such outputs require 
a dominant contribution 
$L_{d}\gg L_K$ from a 
wider disk region dominated by radiation pressure; this  
in turn requires conditions where $\dot m \sim 1$, 
consistent with the 
independent requirement  posed by the high levels  
of $L_{th}$
discussed in \S 2.    

The maximal power expected may now approach $10^{48}$ erg/s, 
exceeding the maximal $L_K$ by some $10^2$ (Livio et al. 1999).
This obtains  if the 
magnetic field in the disk is and remains  
a few times larger than the value threading the hole horizon 
out to distances  $r_e \approx 5 \, r_g$.
Here both $L_d$ and the thermal emission
$L_{th}$ are directly fed by current accretion; 
both scale as  $\dot m$, 
and  $L_{nth}/L_{th}\sim 1$ is expected to hold. 
Now the hole contribution is subdominant, but is likely to 
provide a ``high-velocity spine'' 
crucial for the outward propagation of the jet (see Livio 1999, 
Chiaberge et al. 2000).

The detailed share between $L_d$ and $L_{th}$ depends on 
the balance between two scales 
in the power spectrum of the magnetic field inhomogeneities: 
the large-scale, coherent  vs.  
the small-scale, turbulent 
component. A dominant share of the former, such as to yield a large 
$L_d$ and correspondingly an additional, large  
transfer of angular momentum outwards, may require 
modifications of the standard disk models 
(see Salvati 1997).  

Alternatively, to account for the huge FSRQ outputs one would need
very strong $B$ threading the hole, up to the values $B^2/8\pi\sim \rho c^2$ 
in the plunging orbit region advocated by Meier (1999). 
Such enhanced fields have been argued and discussed variously; 
they look unlikely in   
a thin disk,  whereas in thick disks their status 
is still uncertain. We note they would require high levels of $\dot m$ anyway, 
the main issue here.

In sum, independently of the open if interesting questions concerning 
the detailed modes of power production, 
these considerations strongly suggest a  trend  
toward  weaker powers $L$ but higher ratios $L_{nth}/L_{th}$
 as $\dot m$ is decreased from values 1-10  that mark   
the FSRQs  to $10^{-2}$ that mark the BL Lacs.
We (CD02) referred  to this trend as the  {\it Blazar Main Sequence} (BMS),   
with the implication that it holds   
independently of any genetic link between BL Lacs and FSRQs.

\section{The Blazar Main Sequence}

The BMS may be usefully  contrasted  with the stellar MS. 
Stars were placed,  
even before their energy source was properly understood, 
on the HR diagram relating the simplest observables: 
luminosity and  color, or black body peak $\nu_{b} \propto T_e$.
Now the MS comprises the sources whose luminosity is powered by central 
nuclear burning of H. In stars, the  spectra of the outgoing radiation are 
eventually shaped in the upper envelope 
under conditions still close to optically thick.   
All processes occur in conditions close to hydrostatic and 
thermal equilibrium; so the spectra 
are closely black body, looking 
like spikes at $\nu_b$ on a broad frequency span, while   
$L$ increases sharply 
(and the lifetimes decrease) with increasing $\nu_p$.  
The main underlying 
parameter is mass.

\begin{figure}
\epsfysize=7cm 
\hspace{3.5cm}\epsfbox{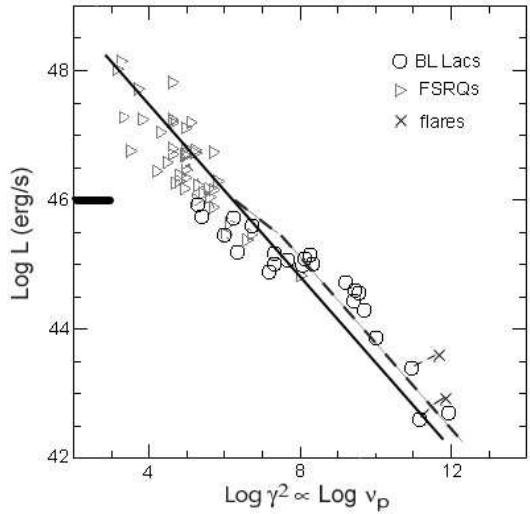} 
\caption[h] {The $L$ vs. $\nu_p$ diagram. Data from Costamante 
et al. (2002). Even the most powerful BL Lacs measured to now fall below $L \sim 10^{46}$ erg/s. 
The solid line represents the trend $L \propto  \nu_p^{-2/3}$ provided 
by eq. (3). The dashed 
line represents the intrinsic scatter predicted in \S 4a. Crosses 
indicate the positions of  Mkn 421  and Mkn 501 at the top of their flares.
The recent data by Giommi et al. (2002a) add substantial scatter particularly in 
the BL Lac section of the diagram.}   
\end{figure}

Blazars, instead, are ruled by {\it strong} gravity; in addition, the 
jet-like, nonthermal component of their radiation is emitted by 
highly {\it relativistic} 
particles under optically thin conditions. 
The lower  peak at $\nu_p$ in their  SEDs is broad, yet sufficiently well 
defined to be represented on the plane labeled by   
the simplest observables: $\nu_p$ itself, and 
the integrated steady luminosity $L$ associated with the peak. Thus a  definite 
``spectral sequence" has been obtained by Fossati et al. (1998), with the 
powerful FSRQs at low  
$\nu_p$ and the weak  BL Lacs at high $\nu_p$;  
the latter data have been recently  
extended by Costamante et al. (2002).
We hold  $\dot m$ to be the {\it main} parameter marking the  BL Lacs from the 
FSRQs (see also B\"ottcher \& Dermer 2002), and underlying the whole BMS. 

We  stress  analogies and differences between the BMS and the stellar MS 
on representing the Blazar  data first in the simplified form of fig. 1. 
The key difference is constituted by the {\it decreasing} trend of $L$ 
vs. $\nu_p$ in the BMS as compared with the increasing trend  in the 
stellar MS;
this BMS feature is resistant to selection effects and 
new detections presented and discussed by Giommi et al. (2002a). 
We trace back such a trend to conditions close to {\it steady} state (steady accretion 
and steady jet outflow) underlying the BMS,   
that replace the equilibrium conditions governing the MS. 

In fact, as a condition for steady energy distribution $N(\gamma)$ 
of the relativistic electrons that emit S and IC radiation,    
one  may equate the acceleration and the radiative 
cooling times, taking up from Ghisellini et al. (1998) and 
B\"ottcher \& Dermer (2002).
So from $\gamma /E \propto 1/\gamma L$   
one obtains the decreasing course $L \propto \nu_p^{-1}$, 
at  {\it given} accelerating  electric field $E$. 

But we find a flatter dependence when  the primary   
$E $ fields are stronger,  due to lower
densities $n \propto \dot m$ in the source.  
For example, consider the  parallel fields 
{\bf E} $\bullet$ {\bf B} $\neq 0$ which arise    
at the flow boundaries where breakdown occurs for  
the force-free approximation governing the bulk of the  
BZ77 magnetospheres; this is bound to occur 
in a space-inhomogeneous and also in a time-depending fashion on short scales. 
Then electrodynamic screening produces  
effective energy gains $E \, d \propto c/\omega_P \propto n^{-1/2}$ that scale 
up when the electron density in the plasma frequency $\omega_P$ is lower; 
so the steady state constraint  $\gamma \, n^{1/2} \propto 1/\gamma L$ 
will allow larger $\gamma$ for a given $L$. 

To see how the course of $L$ vs. $\nu_p$ is modified,     
one may iteratively relate the minimum density to  emissivity to obtain 
$ n \propto L$; matters are simplified on keeping the source volumes 
roughly constant. Now the course flattens to    
$$L \propto  \nu_p^{-2/3} ~,\eqno (3) $$
not a bad rendition even for the overall trend in the data, see fig. 1. 
To a better approximation, one may take into account (see CD02) the full 
dependencies of $\omega_p,\, n $,and  $L/n \propto $ on the 
minimal or maximal electron 
energies $\gamma_{m}, \, \gamma_{M}$ and on the slope of $N(\gamma)$,  
to obtain a sharper  
flattening $L \propto \nu_p^{-1/2}$ in the BL Lac range. Such flatter 
courses may be also viewed at as ``weaker" correlations.

By the same token, we  expect  the maximal electron energies 
to scale up following   
$\gamma_{M} \mincir e\, B\,d/ m_e\, c^2 \sim  10^8 \, B_4\, d_{10}\, 
(M_9/r_{17})^{1.25} \propto L^{-3/4}\div L^{-1}$; 
here the maximum field is expressed as 
$E \mincir 10^4 \, B_4$ G and (following Blandford \& Payne 1982) 
is taken to scale as $ E \propto r^{-1.25}$ 
into the emission region 
at $r \sim 10^{16 \, \pm 1}$ cm, while the effective screening distance is 
$d = 10^{10}\, d_{10}$ cm. So we obtain 
values $\gamma_{M} \sim 10^5$ in FSRQs, and values 
up to $10^2$ times larger in  BL Lacs; these are high enough to 
produce by IC radiation the observed TeV photons. 

We add that  $N(\gamma)$ is expected  
to depart from a pure power-law 
as the acceleration takes place 
in fields endowed with some degree of coherence. In fact, 
we expect the electrons to be accelerated on crossing a sequence of many sheets or 
filaments where the force-free condition 
gradually breaks down on average. 
Correspondingly, the parallel {\bf E } field will progressively 
grow to a point while the width of the filaments or sheets 
widens, with a  corresponding 
growth of the energy gain per acceleration step.  

On the other hand, conditions where the energy gain exceeds  a 
factor 2 (a limiting value in the   
acceleration by relativistic shocks, see Achterberg et al. 2001) to 
approach 10    
are discussed by Massaro (2002). Then   
the energy distribution $N(\gamma)$ has a  
{\it curved} shape, actually a log-parabolic one; the emitted S or IC 
spectra are correspondingly curved over a wide frequency band.    
In fact, the data concerning a number of  BL Lacs  
apparently show such a curvature in the O -- X-ray range (Massaro 2002, 
Giommi et al. 2002b); 
so they call for energy gains of order 10, more appropriate 
for accelerating fields with some degree of 
coherence.

Having outlined the simplified picture, next we point out 
a number of additions yielding variance. 
a) The BMS has an intrinsic width 
related to the  additional parameter $L_K/L_d$ that implies 
within the BL Lac subclass some asymmetric scatter toward high $L$.  
This is because only $L_d$ is directly related 
to  $\dot m$ and to $n$ and (inversely)  to $\nu_p$ 
by the arguments in \S 3, while $L_K$ tends to be
independent when relevant at low $\dot m$;   
then higher non-thermal luminosities $L_{nth} = L_d + L_K$
obtain in BL Lacs at a given $\nu_p$. 
b) We recall that we expect 
$\gamma_{M} \propto  B\, n^{-1/2}$ to hold in the BL Lac range; 
here we add that $B$ is likely to be distributed among the objects with a 
bias toward low values. This will yield lower values of $\nu_p$ and $L$, 
contributing more asymmetric scatter.    
c) Additional scatter toward low $L$ is generally contributed by misaligned sources 
with a given value of $\Gamma$ (see Georganopoulos 2000), and by the 
distribution of $\Gamma$ itself. 

These different scatter components will superpose in the BL Lac range 
to a flattening ``bare" correlation, to blur it mainly toward   
low values of $L$ and $ \nu_{p}$ in a manner 
not unlike that observed and discussed by Giommi et al. (2002a).


Finally, episodes of violent variability are observed to occur 
in the form of flares from hours to weeks, particularly in BL Lacs 
at high photon energies (see Costamante et al. 2002). 
Here the steady state constraint clearly does not apply; 
rather, the basic transient  behavior of the S and IC radiations 
$L \propto \gamma^2 \propto 
\nu_p$ provides a first approximation to the departing {\it branches} 
(see fig. 1) traced by flaring objects on the 
$L - \nu_p$ plane.

\section{Cosmological evolutions}

Along the BMS, we expect longer object lifetimes and {\it slower} 
cosmological evolution to occur.   
Basically this is because conditions of weak activity feeding 
on low levels of   
$\dot m$ can live longer than those of hyperactivity requiring high $\dot m$.   

\begin{figure}
\epsfysize=7cm 
\hspace{3.5cm}\epsfbox{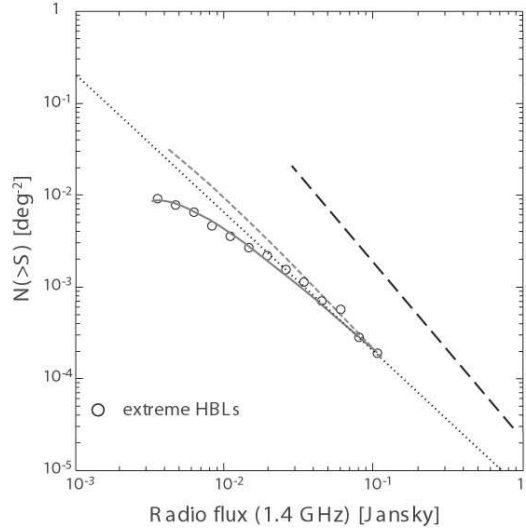} 
\caption[h]%
{The BL Lac counts evaluated 
from the expression given by CD02.
The short-dashed line represents the result from pure LE with time scale 
$\tau_L = 7$ Gyr. The solid line represents the result when  
negative DE with scale $\tau_D = - 5 $ Gyr is included.   
We show also the FSRQ counts from the same expression,  
with $\tau_L =  2$ Gyr and 
$\tau_D = +\, 5$ Gyr. The dotted line represents the 
``Euclidean'' slope.  Hubble constant $H_0 = 65$ km/s Mpc.
Data for BL Lacs from Giommi, Menna, {\&} Padovani (1999).}
\end{figure}  

This may be quantified on considering that roughly similar BH masses of order 
$M \mincir 3 \, 10^9 \, M_{\odot}$ found (see Treves et al. 2002) 
in BL Lacs and in  FSRQs imply a similar 
limit to cumulative masses accreted during the activity of either type.
As to the latter, the mass constraint bounds the cumulative activity time   
to a few  Gyrs with a duty cycle of order $10^{-1}$ 
(Cavaliere \& Padovani 1989). 
 
As to  the former, instead,    
the constraint allows a lifetime up to $10$ Gyr at levels $10^{45}$ erg/s 
or weaker; note that a duration $ E/L_K \sim 5$ Gyr can 
be sustained  just by the rotational energy extractable from  
the coupled system Kerr hole - inner disk, in the absence of  
angular momentum replenishment (Cavaliere \& Malquori 1999). 
Since a similar time behavior $L(t)$ is shared by these objects at 
various power levels, such values apply to the specific 
scale $\tau_L \sim 10$ 
Gyr for the ``luminosity evolution'' of  their population. 

On these simple grounds we expect the space distribution of the 
BL Lacs to be {\it uniform} 
to a first approximation, out to some 
$z \approx 1$;  so we expect from the volume test values  
$\langle V_e/V_a \rangle \approx 0.5$,  
and for the number counts  a nearly Euclidean shape $N(>S) \propto S^{-1.5}$ at 
intermediate fluxes.  
This, in fact, what the data indicate to a first approximation,  
see Giommi, Menna \& Padovani (1999). 

A quantitative prediction is obtained by CD02 on using the expression 
for the counts  at intermediate fluxes in terms of evolutionary times, 
in turn evaluated in more detail.   
The results for BL Lacs counted in the radio  band are  plotted in fig. 2;  
the short-dashed line is obtained  
on using only the basic time  scale  
$\tau_L\approx 7$ Gyr that marks their slow LE.

In contrast (see the long-dashed line in fig. 2), quite steeper counts obtain 
for the FSRQs using  the time scale $\tau_L\approx 2$ Gyr appropriate 
for their strong 
evolution, similar 
to that of radioquiet QSs (Goldschmidt et al. 1999; Giommi, Menna  
\& Padovani 1999). 
The latter behavior is widely  
traced back to exhaustion on the scale of a few Gyrs  
of the gas stockpiled in the host galaxy and usable for accretion; 
this is caused by  previous 
accretion episodes, in addition to   
 ongoing star formation (Cattaneo, Haehnelt \& Rees 1999; 
Haehnelt \& Kauffmann 2000; Cavaliere {\&} Vittorini 2002).   
Exhaustion is bound to occur at $z < 2.5$, when the formation 
of cosmic structures by hierarchical clustering 
evolves beyond the galactic scales;  then 
violent merging events assemble the galaxies into richer and richer 
groups, but rarely reshuffle the galactic masses 
and no longer import much fresh gas supplies into the host galaxy.

In fig. 2, another  time scale scale appears, 
namely, $\tau_D\approx  \pm \, 5 $ Gyr; 
this is negative for the BL Lacs (solid line), meaning 
{\it negative} ``density evolution", that is, birth of these objects at 
moderate or low $z$.
In these conditions, less and less sources are found 
on looking out to increasing $z$; this implies $V_e/V_a < 0.5$ and 
flattening of the counts below the  
Euclidean slope at relatively high fluxes, where the low redshifts 
of the BL Lacs still prevent the cosmological convergence of the volumes 
from operating.     

Negative evolution is not yet apparent in 
all existing the BL Lac surveys 
(see Caccianiga et al. 2001, Perri et al. 2002), but it 
will eventually provide an interesting telltale about the origin of 
these peculiar AGNs.  
To illustrate the issue, let us consider  
a genetic link between the Blazar subclasses
such that BL Lac birthrate $\approx -$ FSRQ deathrate.


First, consider that such a link is a conceivable one, since 
the basic engines are similar while 
the fueling rates are lower for the BL Lacs. So  
these can constitute weaker and later stages  
of earlier and stronger FSRQs, provided  
the cosmic structures evolve so as to 
make less and less mass available for accretion; but this is just what 
has been anticipated  
above on the basis of the QS evolution alone.

Second, the link is  supported by  
a closer consideration of  the strong  evolution shared by the FSRQs 
with the other QSs. For 
example, Cavaliere \& Vittorini (2002, see also refs. therein)
trace back the accretion episodes that feed the 
QSs to triggering interactions of the host galaxies 
with their companions in a group.
These events destabilize the host gas and start  
inflow toward the nucleus; this develops over a galactic dynamical 
time scales of  
a few $10^{-1}$ Gyr, but 
some $5$ similar cycles are expected per host galaxy since 
$z \approx 2.5$. The outcome is exhaustion of the 
host gas on a time scale of 2-3 Gyr. 

Two consequences are expected. For one,  the efficiency of such episodes  
drops on the same  scale; 
so the average QS luminosities are bound to 
halve on the scale $\tau_L \approx 2$ Gyr, to be observed as 
strong LE in the population.
Moreover,  the frequency of such interactions is bound to dwindle   
as the groups evolve into clusters with lower galaxy densities and higher 
velocity dispersions; this 
gives rise to positive DE developing over the longer 
scales $\tau_D \sim $ several Gyrs set by the hierarchical 
clustering. These two evolutionary components 
are recognized in the QSO data, see Boyle et al. (2000).

Thus we expect the powerful FSRQs 
will fade out over times of some $10^{-1}$ Gyr 
after a ``last interaction'' and a last episode of 
accretion of mass with associated  
angular momentum. 
In some 1/2 of the instances this will add constructively to the 
pre-existing {\bf J}, leaving behind a fast  
spinning hole and a long lived BL Lac. 

Thus the scale  for BL Lac births 
will be close to $-\tau_D$, i.e., the negative of that   
 for deaths of bright FSRQs.
In the formers' counts this offsets the slow LE, 
to the effect of  flattening them  
{\it below} the  Euclidean slope as 
illustrated in fig. 2. We add that 
the count normalization, i.e.,  the object number,  
is  bound to be at least a factor 5 below  
the FSRQs' which undergo repeated activity cycles. 

\section{Endpoints}

Note that flattening of the counts is also contributed at faint fluxes 
by the shape of the LF, which is itself flattened at its lower end 
by the beaming effects (see Urry {\&} Padovani 1995);  
but this hardly 
could swamp a  strong evolution of BL Lacs, if it were present.  
In fact, the similarly affected LF of the FSRQs yields counts that 
do show a steep bright section just corresponding to a QS-type intrinsic  
evolution. 

In the clear absence of the latter, there is scope in focusing on   
the negative BL Lac evolution that is now emerging 
from the statistical noise in  
larger and deeper surveys with substantial redshift information (Perri et al. 
2002). If nailed down at the quantitative levels predicted 
above, we stress it will provide statistical evidence that transitions 
FSRQs $\rightarrow$ BL Lacs do occur.  

This will clearly mean a key step toward {\it Grand-Unification} of the 
Blazars. It will also have implications for the radio sources. If 
FR I and FR II constitute the
parent population of the BL Lacs and of the FSRQs, respectively,  
then similar transitions  from II to I --  once a forbidden proposition,  and 
now one more widely entertained, see Padovani \& Urry (2001) and refs. therein --
will be made acceptable or even likely.   

The other, direct evidence toward Grand-Unification may be provided by 
finding at $z \mincir 1$ the {\it transitional} Blazars implied by the 
genetic connection from FSRQs to BL Lacs. We expect that,   
when caught in the act of switching at regimes  $\dot m \sim 10^{-1}$,  
such objects will show not only intermediate  $L$, but also emission lines and 
BBB still shining; when the latter is subtracted from the continuum 
in the manner discussed by D'Elia \& Padovani (2002),  
the residual non-thermal SED ought to be peaked 
at $\nu_p$  beyond the 
optical frequencies. 
Not many of these 
objects are to be expected,  since by definition they will be weaker than the 
canonical FSRQs while the transition will take times 
shorter, if anything, than the lifetime at the top of $\dot m $.
A number of these objects may have already been  found, see Sambruna, Chou 
\& Urry (2000); Perlman et al. (2001); Padovani et al. (2002). 

\bigskip
\centerline{\bf Table 1 - The Blazar Main Sequence }
\begin{table}[h]
\hspace{1.0cm} 
\begin{tabular}{|l|c|c|c|}
\hline
          &FSRQs$^{\phantom{0}}$  &BL Lacs & $\rightarrow$ CR accelerators \\
\hline
optical features      &em. lines, bump$^{\phantom{0}}$& no lines \& bump   &  none    \\

power      &$L \mincir 10^{48}$ erg s$^{-1}$
		&$L \mincir 10^{46}$ erg s$^{-1}$ &
		$L \mincir 10^{42}$ erg s$^{-1}$ \\

evolution             & strong &weak if any &negligible \\		

top energies              & $h\nu \sim 10$  GeV   & $h\nu \sim 10$ TeV&
${\cal E}_{M}\sim 10^{20}$ eV  \\

hole vs. disk           &$L_K \ll L_d$     & $L_{K} \mincir L_d$&
vanishing $L_K$, $L_d$\\ 

key parameter      &$\dot m \sim 1$      &$\dot m \sim 10^{-2}$
   &$\dot m \mincir 10^{-4}$     \\
\hline
\end{tabular}
\end{table}

Waiting for these two lines of evidence to consolidate, 
an eye should be kept open on a further connection with sources 
of high energy cosmic rays. 
This is conceivable and even likely if we carry the BMS and the genetic 
link to their extreme; starting from 
the relation  $\gamma_M \propto L^{-3/4}$ one expects   
 values of order  $10^{10}$ in cases where 
the total output is   $L\mincir 10^{42}$ erg/s. 

So one may envisage {\it  endpoint} objects 
(``dead BL Lacs'') with
very low residual $\dot m < 10^{-3} $ and possibly in full ADAF conditions, 
which would feature very faint if any e.m. emission. But 
they would still support  
strong, nearly unscreened electric fields
under the widely entertained  assumption 
that magnetic fields of order $B \sim 10^2- 10^3$ G can still be 
held by such vestigial disks. So they    
can accelerate particles including protons up 
to limiting energies   
$ {\cal E}_{M}\sim 4\,e\,B\,r_{g}\,(r_{g}/r)^{0.25}\sim 
10^{20}\,M_9^{1.25}\,B_3\,r_{16}^{-0.25}$ eV.  
 Such limiting energies, long recognized to be accessible to  
rotating holes, gratifyingly cover the 
upper range of the ultra-high energy cosmic ray spectrum.

Output 
levels  $L\sim 10^{42}$ erg/s as envisaged above
can just provide the observed UHECR flux;  
tens of these accelerators 
could lie within some 50 Mpc and evade in the simplest way 
the GZK cutoff (Boldt {\&} Ghosh 1999). 
Intergalatic magnetic fields 
of nG strength would blur the geometrical memory of the sources 
for most except for the closest UHECR events. 

We summarize  in Table I the main Blazar features that the BMS can explain 
or predict. In fig. 3  we 
schematically illustrate  the further steps warranted 
by negative evolution and transitional objects: 
Grand-Unification of Blazars, and the link with 
UHECR accelerators. 



\begin{figure}
\epsfysize=5cm 
\hspace{2cm}\epsfbox{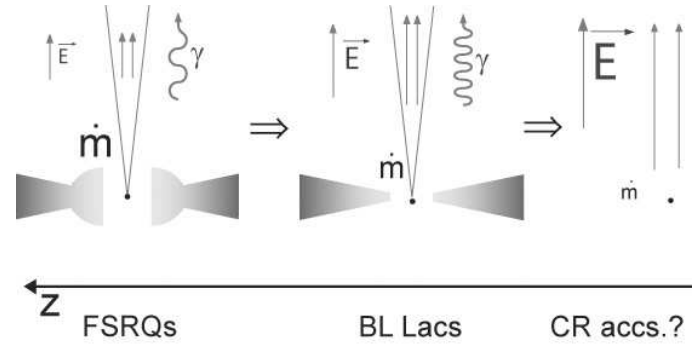} 
\caption[h]{To illustrate the BMS, and the possible genetic connection 
from the hyperactive FSRQs at high-medium $z$ to the moderate or weak, 
low $z$ BL Lacs. This may be extended  
down to  $z\approx 0$ to including as  local relics the objects, 
e.m. silent but still 
effective as particle  accelerators, discussed in \S 6}  
\end{figure}

\newpage

\end{document}